\documentstyle[12pt,aasms]{article}
\begin{document}
\newcommand{\cm}{$\rm cm^{-3}$}
\newcommand{\ergs}{$\rm ergs \, kpc^{-2} \, s^{-1}$\ }
\newcommand{\kms}{$\rm km \, s^{-1}$}
\def\etal{{\it et~al.\ }}
\def\eg{{\it e.~g.\ }}
\def\ie{{\it i.~e.,\ }}

\title{Superbubble evolution including the star-forming clouds: 
Is it possible to reconcile LMC observations with model predictions?}

\author{Sergei Silich\altaffilmark{1} and Jos\'e Franco\altaffilmark{2}}

\altaffiltext{1}{Main Astronomical Observatory National Academy of Sciences
of Ukraine, 252650 Kyiv, Golosiiv, Ukraine\\
(E-mail: silich@mao.kiev.ua)}

\altaffiltext{2}{Instituto de Astronom\'{\i}a-UNAM, Apdo. Postal
70-264, 04510 M\'exico D. F., Mexico\\
(E-mail: pepe@astroscu.unam.mx)}

\begin{abstract}
Here we present a possible solution to the apparent discrepancy between the 
observed properties of LMC bubbles and the standard, constant density bubble 
model. A two-dimensional model of a wind-driven bubble expanding from a 
flattened giant molecular cloud is examined. We conclude that the expansion 
velocities derived from spherically symmetric models are not always
applicable to elongated young bubbles seen almost face-on due to the
LMC orientation. In addition, an observational test to differentiate
between spherical and elongated bubbles seen face-on is discussed. 

\keywords{Galaxies: Kinematics and Dynamics --- ISM: Bubbles --- ISM: Clouds
---  ISM: Kinematics and Dynamics ---  ISM; Shock Waves  --- ISM: Structure
--- Magellanic Clouds}

\end{abstract}

\section{Introduction} 

Since the discovery of large shells and holes of neutral hydrogen in the 
Magellanic Clouds (see McGee \& Milton 1966 and Westerlund \& Mathewson 1966 
for the LMC, and Hindman 1967 for the SMC), the Milky Way (Heiles 1979), and 
M31 (Brinks \& Bajaja 1986), the study of interstellar bubbles has been 
extended to several nearby galaxies (see recent reviews by Brinks \& Walter 
1998, and Thilker 1998). The optical counterpart of these objects are the 
H$_{\alpha}$ ring-shaped nebulae (\eg Boulesteix \etal 1974; Davies
\etal 1976; Sivan 1977; Pellet \etal 1978; Meaburn 1980; Lozinskaya
\& Sitnik 1988), which are powered by young massive stars, and many
HI bubbles are actually delineated by them. Thus a paradigm for
bubble evolution driven by energy injection from massive stars was
developed during the sixties and seventies by Pikel'ner (1968),
Avedisova (1972), and Weaver \etal (1977). These original analytical 
models have been extended during the last two decades with 2D and 3D
numerical simulations by Bisnovatyi-Kogan \& Blinnikov (1982), Mac
Low \& McCray (1987), Palous (1992), and Silich (1992); see reviews
by Tenorio-Tagle \& Bodenheimer (1988) and Bisnovatyi-Kogan \& Silich
(1995). 

A number of important processes affecting the expansion of shells have been 
studied during this time, including the effects of blowout and dynamical 
instabilities in decreasing density gradients (Mac Low \etal 1989; 
Tenorio-Tagle \etal 1990; Garc\'{\i}a-Segura \& McLow 1995a, 1995b), 
gravitational instabilities (McCray \& Kafatos 1987; Ehlerova \etal 1997), 
ambient magnetic fields (Tomisaka 1990, 1998; Ferriere \etal 1991), galactic 
differential rotation (Palous 1992; Silich \etal 1996; Moreno \etal 1999), 
radiation pressure from field stars (Elmegreen \& Chiang 1982; Franco \etal 
1991), the impact of supernova fragments in expanding superbubble shells 
(Franco \etal 1993), the role of hydrodynamic ablation and thermal
evaporation of ambient clouds (Hartquist \etal 1986; Arthur \& Henney
1996; Silich \etal 1996), and photoionization from the central stars
(Comeron 1997). 

In principle, then, one could compare the predictions of a variety of
different models (\ie the resulting bubble shapes, expansion
velocities, column densities, X-ray luminosities, etc.) with the
available observational data for holes and shells (\eg Chu \& Mac Low
1990; Silich \etal 1996; Mashchenko \etal 1998; Thilker \etal 1998).
A direct comparison with observations, however, presents several
problems because it is difficult to constraint most of the relevant
model parameters (for instance, the energy input rate and the
original density structure). Nonetheless, several steps have been
done recently to overcome some of these questions, and additional
problems with the applicability of models have emerged.

Detailed studies of the OB stellar content in associations in the Milky Way 
and the Magellanic Clouds are now possible with CCD photometry (see
Saken \etal 1992; Oey 1996a), and, in combination with stellar
evolution models, they provide limits to the mechanical energy input
rate. These rates have been used to compare LMC bubble observations
with the predictions of the ``standard'' model (\ie a spherically
symmetric shell evolving in a constant density medium). The
comparisons indicate that the standard model cannot reproduce the 
properties of a number of well observed cases (Rosado 1986; Oey \&
Massey 1995; Oey 1996b). In particular, the collection of bubbles
observed in the LMC exhibit two different sets of objects with
conflicting size-velocity relations. Given that the observed shell
sizes are well known, the problem can be reduced to the existence of
objects with expansion velocities that are either too high 
or too low to be explained by the simple standard model. For the low
velocity objects, the discrepancy could probably be explained by
errors in the estimation of either the input wind power or ambient
gas density (Oey 1996b). For the high velocity objects, however, the
observed nebular expansion velocities ($V_{exp} \ge 25$ km s$^{-1}$)
are at least a factor of two larger than the expected values (Oey
1996b). Neither the density gradient in the disk of the LMC nor
possible variations of the initial mass function (or non-coeval 
star formation) can resolve this discrepancy (Oey 1996b).  

A possible solution to this problem can be associated with the fact that the 
LMC has a moderate inclination angle, of about 27$^{\circ}$ (Crampton 1979),
and the shell are viewed with a nearly face-on orientation. For face-on 
galaxies, the density gradient along the $z$-direction can influence the bubble 
expansion along the line of sight, increasing the observed gas velocities. The 
effects of the gradient are more relevant in big spirals with a thin disk, but 
are less important in dwarf irregulars with extended HI layers (see discussion 
by Brinks \& Walter 1998). However, the overall effect becomes certainly 
pronounced, even for dwarfs with very thick layers of H~I, if one takes into 
account the presence of the parent giant molecular cloud (GMC) which gives 
birth to the perturbing stellar group and controls the initial bubble 
expansion: the high mass concentration within the parent GMC induces strong 
changes in the dynamics of shocks and can accelerate and generate fragmentation 
in the resulting shells (see Franco \etal 1989, 1990, 1997 and
Garc\'{\i}a-Segura \& Franco 1996). A recent semi-analytical study of a model 
with a sharp density contrast indicates that the bubble kinematics could reach 
the required velocity values (Oey \& Smedley 1998), and the presence of the 
parent cloud provides the required density gradient. In this paper we focus on 
the dynamics and observational manifestations of relatively young bubbles, with
modest sizes (below 100 pc), which originate from associations embedded inside 
GMCs. The paper is organized as follows. Section 2 describes the basic input 
model and numerical scheme. Section 3 contains the results from numerical 
calculations, and these results are discussed in section 4.

\section{The cloud density structure and model assumptions}

For a spherically symmetric isothermal self-gravitating cloud in equilibrium, 
the gas density declines as $r^{-2}$ (where $r$ is the distance from the cloud
center). For cylindrical (disk-like) self-gravitating clouds with infinite 
radius, on the other hand, the isothermal density stratification along the 
$z$-axis varies as sech$^2 (z/H)$ (where $H$ is the scale height). Obviously, 
other cloud models with different morphologies result in different functional 
forms for the density stratifications. Here we use a simplified model to 
simulate the density distribution of a flattened, two-dimensional GMC. The 2-D 
stratification is defined in the cylindrical coordinate system $(r, \phi , z)$,
with the origin at the cloud center. The cloud has a constant density core, 
with density $\rho_c$, and the density decreases as a power-law until it 
reaches the value of the ambient medium, $\rho_{ISM}$. For simplicity, here we
assume a constant value for $\rho_{ISM}$, and the initial GMC density 
distribution is then defined as
%-------------------------------------------------------------------- 
\begin{equation}
      \label{eq.1}
\rho  = \left\{
\begin{array}{lcl}
\rho_c, \hfill \ \ {\rm for} \ \ \left(\frac{r}{R_c}\right)^2 +
\left(\frac{z}{Z_c}\right)^2 
                 \quad \le  1, 
\\ [0.2cm]
 \rho_c \left[\left(\frac{r}{R_c}\right)^2 + \left(\frac{z}{Z_c}\right)^2
        \right]^{-w/2},
 \hfill \ \ {\rm for} \ \ \left(\frac{r}{R_c}\right)^2 +
\left(\frac{z}{Z_c}\right)^2 
        \le \xi^{2/w},
\\ [0.2cm]
\rho_{ISM}, \hfill \ \   {\rm for} \ \  
\left(\frac{r}{R_c}\right)^2 + \left(\frac{z}{Z_c}\right)^2
        > \xi^{2/w}, \\
\end{array}
\right.
\end{equation}
%-----------------------------------------------------
where $\xi = \rho_c / \rho_{ISM}$ is the ratio of the cloud core density to the 
ISM gas density, $w$ is the power-law index, and $R_c$ and $Z_c$ are the 
characteristic scale heights for the cloud density distribution in the $r$ and 
$z$-directions, respectively. The resulting maximum cloud extent along any of 
these axes is defined by $r_{cl} = R_c \xi^{1/w}$ and $z_{cl} = Z_c \xi^{1/w}$.

The appropriate range of values for the cloud parameters can be derived from
observational results. For instance, using the spherically symmetric case ($R_c 
=Z_c$) we can derive the core radius from the condition that the observed shell
mass, M$_{obs}$, is contained within the observed shell radius, $R_{obs}$. For 
a given core density,  the observed mass is 
simply given by
%----------------------------------------------------------------
\begin{equation}
      \label{eq.2}
M_{obs} = M_c + 4 \pi \int_{R_c}^{R_{obs}} \rho(r) r^2 {\rm d}r,
\end{equation}
%---------------------------------------------------------------------
where $M_c$ is the core mass,
%For the modest shell sizes that we consider in this paper, $r_{cl} > R_{obs}$, 
and the resulting core radius follows from the equation 
%----------------------------------------------------------------
\begin{equation}
      \label{eq.3}
\left(\frac{R_c}{R_{obs}}\right)^3 - \frac{3}{w} \left(\frac{R_c}{R_{obs}} 
\right)^w + \frac{3-w}{w} \frac {3 M_{obs}}{4 \pi R_{obs}^3 \rho_c} = 0.
\end{equation}
%---------------------------------------------------------------------
For simplicity, we use $w=2$ (which corresponds to a self-gravitating and
isothermal sphere in the $R_c=Z_c$ case), and the radius $R_{obs}$ is set 
equal to 40 pc. The mass M$_{obs}$ is considered in the range 2 to $5\times 
10^4$ M$_{\odot}$, to be consistent with the observed shell masses
(Oey 1996b). The solution of equation (\ref{eq.3}), which is solved
numerically at the beginning of the runs, is used as the
characteristic scale height $R_c$ for the two-dimensional models, and
the characteristic scale in the $z$-direction is reduced to a half of
this value, $Z_c=R_c/2$. Finally, we take the cloud core number
density, $n_c$, as a free parameter, and explore density values in
the range from 10 to 10$^2$ cm$^{-3}$. We designate the models as 'A'
or 'B', depending on whether the resulting cloud mass $M_{obs}$ is
equal to 2 or $5\times 10^4$ M$_{\odot}$, respectively. Thus, for a
given core density value, they correspond to small and large clouds,
respectively. The resulting GMC density distributions for both types
of models, A and B, are illustrated in Figure 1, where the last
isodensity contour represents the cloud boundary. 

Assuming that the initial GMCs are in hydrostatic equilibrium,
self-gravity defines the total pressure at the cloud center. Again,
using the spherically symmetric model as an illustrative case, the
total pressure at the cloud center is (Garc\'{\i}a-Segura \& Franco
1996) 
%----------------------------------------------------------------
\begin{equation}
      \label{eq.6}
p_c = \frac{16 \pi}{9} G \rho_c^2 R_c^2 . 
\end{equation}
%---------------------------------------------------------------------
The thermal pressure inside an adiabatic bubble, on the other hand, is 
%----------------------------------------------------------------
\begin{equation}
      \label{eq.7}
p_{in} (r_s) = \frac{ (4.14 \rho)^{1/3} \ L_0^{2/3}}{2 \pi r_s^{4/3}},
\end{equation}
%---------------------------------------------------------------------
where $r_s$ is the bubble radius, and $L_0$ is the mechanical energy input 
rate. The thermal pressure inside a bubble with radius equal to $R_c$ (\ie 
a supershell emerging from the cloud core) exceeds the total pressure at the 
cloud center, $p_{in} (r_s=R_c) > p_c$, when the core densities are
below the reference value
%----------------------------------------------------------------
\begin{equation}
      \label{eq.8}
n_{\rm ref} =  \left(\frac{1}{4.14}\right)^{1/3} \left(\frac{9}{32 \pi ^2 
G}\right)^{3/5} \left(\frac{L_0}{R_c^5}\right)^{2/5} \simeq 2 \times 10^4 \ 
L_{36}^{2/5} \left(\frac{\rm 1 \ pc}{R_c}\right)^2 \ {\rm cm}^{-3},
\end{equation}
%---------------------------------------------------------------------
where $L_{36} =L_0/10^{36}$ erg s$^{-1}$. Thus, for densities below $n_{\rm 
ref}$, one can neglect the cloud gravity and pressure during the early 
expansion and, for simplicity, we maintain only the external ISM pressure, 
$p_{ISM} = k \, n_{ISM} \, T_{ISM}$, where $k$ is the Boltzmann's constant (the 
ambient gas temperature is maintained at $6 \times 10^3$ K$^{\circ}$ throughout 
the calculations). 

The simulations are performed with the three-dimensional code described by Bisnovatyi-Kogan \& Silich (1995) and Silich \etal (1996), which is based on 
the thin layer approximation. In the present set of calculations the energy 
input rate is assumed constant during the runs. Shell evaporation into the 
bubble due to thermal conduction is taken into account (see Silich \etal 
1996), and this is the only source for mass injection into the cavity. The 
calculations of the X-ray luminosities are done with the table for the specific 
X-ray emissivities described by Suchkov \etal (1994). Several runs had been 
done taking into account possible fragmentation of the shell via the
Rayleigh-Taylor (R-T) instability, as discussed by Silich \& Tenorio-Tagle 
(1998). The model parameters used in the runs are summarized in Table 1.

\section{The results}
  
Figure 2 shows the resulting morphologies for models A and B (in the left and
right panels, respectively). As expected, after expanding inside the constant 
density core, the remnant is elongated along the $z$-axis, where the density 
gradient is steepest. The deviation from the spherical morphology is already 
apparent after 2 Myr of evolution (Figs. 2b and 2e). At late
evolutionary times, a dense, compressed ring-like belt is formed at 
the midplane of the cloud, as can be noticed in Fig. 2c (at 4 Myr)
for model A2. In fact, model A2 presents a well defined hour-glass
shape after 4 Myr, with two semi-spheres separated at midplane, as
described in the analytical approach of Kontorovich \& Pimenov 
(1997). Model B2, on the other hand, evolves more slowly in the
higher density cloud, and after 4 Myr (Fig. 2f) it looks similar to
model A2 after 2 Myr (Fig. 2b). Despite the large differences in the
$z$-direction between the elongated and spherical models, the
midplane radii are similar for both types of models. 

The evolutionary tracks for the two sets of cloud models are shown in
Figures 3 and 4 (models A and B, respectively). For comparison,
panels 3a and 4a show the radii and expansion velocities (solid and
dotted lines, respectively) for the corresponding spherical bubble
cases. Panels 3b, 3e, 4b, and 4e illustrate the kinematics for
elongated bubbles, as they should be seen in a face-on galaxy: 
the solid lines represent the bubble radii in the midplane of the
host galaxy (\ie along the $r$-axis), whereas the dotted lines
correspond to the shell velocity in the $z$-direction. The
kinematical properties for edge-on galaxies are shown in panels 3c,
3f, 4c, and 4f. In this case, the dotted lines represent the
expansion velocities in the galactic midplane, and the solid and 
dashed lines show the bubble semi-axes along the plane and in the 
$z$-direction, respectively. For completeness, panels 3d and 4d show two runs 
similar to the ones displayed in 3b and 4b for the face-on configuration but 
allowing for shell fragmentation via the R-T instability. The evolutionary 
tracks are similar, except for a small increase in the expansion
velocities for the cases allowing for the R-T instability. 

The results presented in Figures 3 and 4 indicate that the expected
expansion velocities for elongated bubbles in face-on galaxies are
certainly larger than those derived from spherically symmetric models.
The corresponding radii, however, are almost identical in both cases. A
comparison between the results for small and large clouds (models A
and B, respectively) show that the departures from the spherical case
have well defined trends. For instance, the maximum value of the
expansion velocity for smaller clouds is reached much 
earlier than in the case of larger clouds. Also, the velocity tracks for 
bubbles generated from small clouds are more peaked and can reach
higher velocity values. Thus, as the cloud size increases, the peak
velocity value decreases and the evolutionary track becomes
shallower. In all flattened cloud cases, however, the expansion
velocity remains well above the one of the spherical model. For the
particular cases that we show here, the expansion velocities can be
higher than 20 km s$^{-1}$ during the first 3 Myr of evolution. 

These results raise the obvious question of how can one distinguish between 
spherical and elongated objects in a face-on galaxy. The simplest way to 
resolve this issue is to look for differences in the velocity distribution 
as a function of the bubble radius. To make the results independent of any 
particular model, we define the impact parameter $a$ as the
normalized distance from the bubble center. The normalization is done
with the maximum shell radius, $a=r/R_{max}$, for the face-on
configurations, and with the maximum $z$-extension, $a=z/Z_{top}$,
for the edge-on cases. The $z$- and $r$-components of the expansion
velocities are also normalized to the maximum projection velocity,
$U_{max}$. The resulting velocity distributions for models 
A are shown in Figure 5 (the results for models B are qualitatively similar). 
The solid lines represent the spherical model A1, and the dotted and dashed 
lines correspond to the face-on and edge-on configurations of model A2, 
respectively. Panel 5a displays the velocity distribution after $2
\times 10^5$ yr, and panel 5b shows the same distribution after 1
Myr. At this time the elongated shell has an hour-glass form already,
but the distributions for spherically symmetric and elongated face-on
objects remain similar. A clear difference appears in the maximum
value for the edge-on objects that shifts to the locations away from
the center (dashed lines). The differences among the three cases are
apparent only after 2 Myr, and the corresponding velocity 
distributions show distinctive features. The spherical case maintains its 
initial monotonic form and the edge-on case remains peaked
off-center, but now the face-on case becomes double valued in regions
close to the edge of the shell. This second velocity component near
the shell boundaries becomes the kinematic feature that can help to
recognize the existence of elongated bubbles seen face-on. For
completness, Figure 6 displays isovelocity contours for model 
A2 at 1 and 2 Myr of evolution, and as seen in face-on and edge-on galaxies.

Finally, Figure 7 illustrates the evolution of the X-ray luminosity,
$L_x$, for the three models A. The bubble densities and temperatures
drop faster in the elongated cases, and the resulting luminosities
have a more rapid drop-off (also, the smaller thermal evaporation
rates at the top shell regions prevent increase in the $L_x$ values).
The X-ray luminosity is stabilized as soon as the bubbles begin to
expand in the external ISM. In contrast, the emission from 
the spherically symmetric case increases monotonically during the first 4 Myr 
of evolution. These differences are due to the differences in kinematical 
evolution, but the present results should considered only lower
limits because there are other important effects that have not been
included here. For instance, the peak $L_x$ values can be increased
if one includes clumps (\eg Arthur \& Henney 1996; Silich \etal 1996)
inside the hot remnant (the form of the curve, however, remains the
same). Also, if one includes the presence of fragments in the stellar
ejecta (Franco \etal 1993), the X-ray luminosity of the expanding shell
is increased at each impact, and the light curve is modified
accordingly. These issues require a more detailed study and will be 
explored in the near future.

\section{Discussion and conclusions}

In this paper we have presented a possible solution to the apparent discrepancy 
between the observed properties of some well observed LMC shells with 
high-velocities and the standard bubble model. Our present model assumes that 
these shells are driven by the energy injection from massive OB stars. This is 
a reasonable assumption for small bubbles, but may not be true for all observed 
holes and shells. In particular, the largest and most energetic observed 
superbubbles could be ascribed to a different, non-stellar, origin (such as the 
collision of high-velocity clouds with the gaseuos disk; \eg Tenorio-Tagle 
\etal 1987; Santill\'an \etal 1999).

The issue at hand is the role of the parent GMC in the kinematical properties 
of young bubbles when viewed, as in the case of the LMC, at an almost face-on 
orientation. For simplicity, our 2-D numerical calculations consider the
presence of flattened GMCs but do not include the $z$-gradient of the main 
gaseous disk. Nonetheless, the results already indicate that bubbles blowing 
out of these flattened clouds can reach a high degree of asymmetry on a short
timescale (during the first million years of expansion), with $z$-velocities in 
the range of the observed high-velocity cases. This is in line with the 
semi-analytical results for a sharp density contrast discussed by Oey \& 
Smedley (1998), but a steep density gradient is not really required. We find 
that for moderate values of the GMC flatness, the expansion velocities at the
top could easily exceed those expected from spherical models by a factor of two 
or more. This scheme then provides a possible explanation of the observed 
high-velocity cases, with $V_{exp} \ge 25$ km s$^{-1}$ and radii of several 
tens of parsecs (Rosado \etal 1981, 1982; Rosado 1986; Oey 1996b).

With respect to the resulting X-ray luminosities, the present version of the 
model cannot explain the X-ray excess that is often observed in high-velocity 
superbubbles (Chu \& Mac Low 1990; Rosado \etal 1993). To explore this issue 
one requires to consider the destruction of pre-existing gas clumps
(Arthur \& Henney 1996; Silich \etal 1996), or to include the
interaction of fragmented ejecta with the expanding shell (Franco
\etal 1993). Each of these mechanisms can increase the value of $L_x$
during different moments of the evolution. 

The remarkable difference in the velocity distributions as a function of 
impact parameter for face-on and edge-on galaxies, illustrated in
Figure 5, indicates that one can differentiate elongated and
spherical shell in both face-on and edge-on galaxies. Spectral line
splitting at the periphery of nebular shells and off-center peak
velocity values are indicators of elongated morphologies in the
face-on and edge-on cases, respectively. Observational studies with
adequate spatial resolution are needed to verify these predictions 
in nearby galaxies with different orientations.

This is the same type of model that is commonly applied to explain fast 
starburst-driven outflows in external galaxies (see review by Heckman 1997),
and we have only added the expected density structure at the initial stages of
outflow evolution. Our results indicate that the presence of star-forming 
clouds produce assymetries in short time scales, and shell projection effects 
are important when comparing models with observed shells. Future studies 
including the destructive effects of expanding HII regions (\eg Franco \etal 
1994), and the re-acceleration generated by individual supernova explosions 
(\eg Tenorio-Tagle \etal 1991; Franco \etal 1991; Arthur \& Falle
1993; Arthur \& Henney 1996), will certainly improve our
understanding of the early phases of superbubble evolution. 

{\bf Acknowledgments} It is a big pleasure to thank Sally Oey and Margarita
Rosado for helpful discussions on LMC bubbles and shells, and Jane Arthur for
useful comments. We also thank Elias Brinks, David Thilker, and Rene
Walterbos for information on superbubble population of spirals and
irregulars. Special thanks to Steve Shore and an anonymous referee
for very useful suggestions that greatly improved the final version
of this paper. JF acknowledges partial support from DGAPA-UNAM grant
IN130698, CONACyT grants 400354-5-4843E and 400354-5-0639PE, and a
R\&D CRAY Research grant. SAS acknowledges support from a Royal
Society grant for joint projects with the former Soviet Union States,
and the staff of IoA and RGO in Cambridge, where this study has been
initiated. He thanks the Instituto de Astronom\'{\i}a-UNAM for
support, hospitality, and friendly assistance during his visit to Mexico. 

{\center \section*{References}} \setlength{\parindent}{-1.0\parindent}

\begin{description}

\item Arthur, J. \& Falle, S. A. E. G. 1993, MNRAS, 341, L63

\item Arthur, J. \& Henney, W. 1996, ApJ, 457, 752

\item Avedisova, V. S. 1972, SovAstron, 15, 708

\item Bisnovatyi-Kogan, G.S. \& Blinnikov, S.I. 1982, SovAstron, 26, 530

\item Bisnovatyi-Kogan, G.S. \& Silich, S.A. 1995, RevModPhys, 67, 661

\item Boulesteix, J., Courtes, G., Laval, A., Monnet, G. \& Petit, A. 1974,
A\&A, 37, 33

\item Brinks, E. 1994, in Violent Star Formation: from 30 Doradus to QSOs,
ed. G. Tenorio-Tagle (Cambridge: Cambridge Univ. Press), 145  

\item Brinks, E. \& Bajaja, E. 1986, A\&A 169, 14

\item Brinks, E. \& Walter, F. 1998, in The Magellanic Clouds and Other 
Dwarf Galaxies, ed. T. Richtler \& J. Braun (Aachen: Shaken Verlag), in press

\item Chu, Y.-H. \& Mac Low, M.-M. 1990, ApJ, 365, 510

\item Comeron, F. 1997, A\&A, 326, 1195

\item Crampton, D. 1979, ApJ, 230, 717

\item Davies, R. D., Elliot, K. H. \& Meaburn, J. 1976, MNRAS, 81, 89

\item Ehlerova S., Palous, J., Theis, Ch. \& Hensler, G. 1997, A\&A, 328, 121

\item Elmegreen, B. G. \& Chiang, W. H. 1982, ApJ, 253, 666

\item Ferriere, K., Mac Low, M.-M. \& Zweibel, E. 1991, ApJ, 375, 239

\item Franco, J., Ferrara, A., R\'o\.zyczka, M., Tenorio-Tagle, G. \& Cox,
D. P. 1993, ApJ, 407, 100

\item Franco, J., Ferrini, F., Ferrara, A. \& Barsella, B. 1991, ApJ, 366, 443

\item Franco, J., Plewa, T. \& Garcia-Segura, G. 1997, in Starburst 
Activity in Galaxies, ed. J. Franco, R. Terlevich \& A. Serrano,
RevMexAA Conf. Series, 6, 172

\item Franco, J., Shore, S. N. \& Tenorio-Tagle, G. 1994, ApJ, 436, 795

\item Franco, J., Tenorio-Tagle, G. \& Bodenheimer, P. 1989, RevMexAA, 18,
65 

\item Franco, J., Tenorio-Tagle, G. \& Bodenheimer, P. 1990, ApJ, 349, 126

\item Franco, J., Tenorio-Tagle, G., Bodenheimer, P. \& R\'o\.zyczka, M. 1991,
PASP, 103, 803

\item Garc\'{\i}a-Segura, G. \& Franco, J. 1996, ApJ, 469, 171

\item Garc\'{\i}a-Segura, G. \& Mac Low, M.-M. 1995a, ApJ, 455, 145

\item Garc\'{\i}a-Segura, G. \& Mac Low, M.-M. 1995b, ApJ, 455, 160

\item Hartquist, T.W., Dyson, J.E., Pettini, M. \& Smith, L.J. 1986, MNRAS, 
221, 715

\item Heckman, T. 1997, in Starburst Activity in Galaxies, ed. J. Franco, R.
Terlevich \& A. Serrano, RevMexAAp (Conf. Series), 6, 156

\item Heiles, C. 1979, ApJ, 229, 533.

\item Hindman, J. V. 1967, AustJPhys, 20, 147

%\item Jones, T. W., Kang, H. \& Tregillis, I. L. 1994, ApJ, 432, 194

\item Kontorovich, V.M. \& Pimenov, S.F. 1997, SolPhys, 172, 93

\item Lozinskaya, T. A. \& Sitnik, T. G. 1988, AstronZhLett, 14, 240

\item Mac Low, M.-M. \& McCray, R. 1988, ApJ, 324, 776

\item Mac Low, M.-M., McCray, R. \& Norman, M. L. 1989, ApJ, 337, 141

\item Mashchenko, S., Thilker, D. \& Braun, R. 1999, A\&A, 343, 352

\item Meaburn, J. 1980, MNRAS, 192, 365

\item McGee, R. X. \& Milton, J. A. 1966, AustJPhys, 19, 343

\item Moreno, E., Alfaro, E. J. \& Franco, J. 1999, ApJ, in press

\item Oey, M. S. \& Massey, P. 1995, ApJ, 452, 210

\item Oey, M. S. 1996a, ApJ, 465, 231

\item Oey, M. S. 1996b, ApJ, 467, 666

\item Oey, M. S. \& Smedley, S. A. 1998, AJ, 116, 1263

\item Palou\v{s}, J. 1992, in Evolution of Interstellar Matter and
Dynamics of Galaxies, ed. J. Palous, W.B. Burton \& P.O. Lindblad
(Cambridge: Cambridge Univ. Press), 65

\item Pellet, A., Astier, N., Viale, A., Courtes, G., Maucherat, A., Monnet, G.
\& Simien, F. 1978, A\&ASuppl, 33, 439

\item Pikel'ner, S. B. 1968, ApLett, 2, 97

\item Rosado, M. 1986, A\&A, 160, 211

\item Rosado, M., Georgelin, Y. P., Georgelin, Y. M., Laval, A. \& Monnet, G.
1981, A\&A, 97, 342

\item Rosado, M., Georgelin, Y. M., Georgelin, Y. P., Laval, A. \& Monnet, G.
1982, A\&A, 115, 61

\item Rosado, M., Le Coarer, E., Laval, A. \& Georgelin, Y. P. 1993, RevMexAA, 
27, 41

\item Saken, J.M., Shull, J.M., Garmany,C.D., Nichols-Bohlin, J. \& Fesen, R.A. 
1992, ApJ, 397, 537 

\item Santill\'an, A., Franco, J., Martos, M. \& Kim, J. 1999, ApJ, in press

\item Silich, S. A. 1992, Ap\&SS, 195, 317

\item Silich, S. A., Franco, J., Palous, J \& Tenorio-Tagle, G. 1996, ApJ, 468, 
722

\item Silich, S. A. \& Tenorio-Tagle, G. 1998, MNRAS, 299, 249

\item Sivan, J. P. 1977, PhD Thesis, Univ. of Provence, France

\item Suchkov, A. A., Balsara, D. S., Heckman, T. M. \& Leitherer, C. 1994, 
ApJ, 430, 511

\item Tenorio-Tagle, G. \& Bodenheimer, P. 1988, ARAA, 26, 145

\item Tenorio-Tagle, G., Franco, J., R\'o\.zyczka, M. \& Bodenheimer, P. 1987, 
A\&A, 179, 219

\item Tenorio-Tagle, G., R\'o\.zyczka, M. \& Bodenheimer, P. 1990, A\&A, 237, 
207

\item Tenorio-Tagle, G., R\'o\.zyczka, M., Franco, J. \& Bodenheimer, P. 1991,
MNRAS, 251, 318

\item Thilker, D.A., Braun, R. \& Walterbos, R. 1998, A\&A, 332, 429

\item Thilker, D.A. 1998, in Interstellar Turbulence, ed. J. Franco \& A. 
Carrami\~nana (Cambridge: Cambridge Univ. Press) (in press)

\item Tomisaka, K. 1990, ApJ, 361, L5

\item Tomisaka, K. 1998, MNRAS, 298, 797

\item Weaver, R., McCray, R., Castor, J., Shapiro, P. \& Moore, R. 1977, ApJ, 
218, 377

\item Westerlund, B. E. \& Mathewson, D. S. 1966, MNRAS, 131, 371

\end{description}
%\end{thebibliography}

\clearpage

%-------------------------------------------------------------
%\small

\centerline {\bf {Table 1}}
\centerline {Model parameters (with $w=2$)}
%\scriptsize
\vspace{0.2cm}
\begin{center}
\begin{tabular} {lccccccc} \hline
Model & n$_c$ & n$_{ISM}$ & M$_f$ & R$_c$ & Z$_c$ & R$_{cl}$ & L$_{OB}$ \\
& cm$^{-3}$ &  cm$^{-3}$ & 10$^4 M_{\odot}$ & pc & pc & pc & 10$^{36}$ 
erg s$^{-1}$ \\
\hline
A1 & 2.4 & 2.4 & 2 & -  & -  & - & 5  \\
A2 & 10  & 0.2 & 2 & 12.6  & 6.3 & 89.5  & 5  \\
A3 & 100 & 0.2 & 2 &  3.7  & 1.8 &  82.1  & 5  \\
B1 & 5.9 & 5.9 & 5 & -  & -  & - & 5  \\
B2 & 10  & 0.2 & 5 & 22.5  & 11.25 & 159.0 & 5  \\
B3 & 100 & 0.2 & 5 &  5.9  &  2.95 & 132.5 & 5  \\
\hline
\end{tabular}
\end{center}
%----------------------------------------------------------------------

\clearpage
%---------------------------------------------------------------------
\begin{figure}
\centerline{\plotone{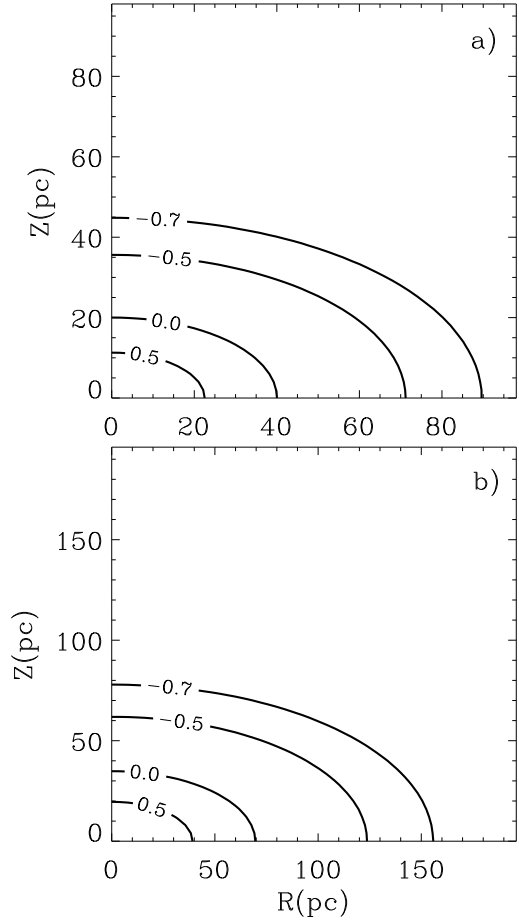}}
\caption{The gas density distribution for the model GMCs. Panels (a) 
and (b) show the isodensity contours for models A2 and B2,
respectively. The value of $\log{n}$ is indicated at each contour.}
\end{figure}
%---------------------------------------------------------------------
\clearpage
\begin{figure}
\centerline{\plotone{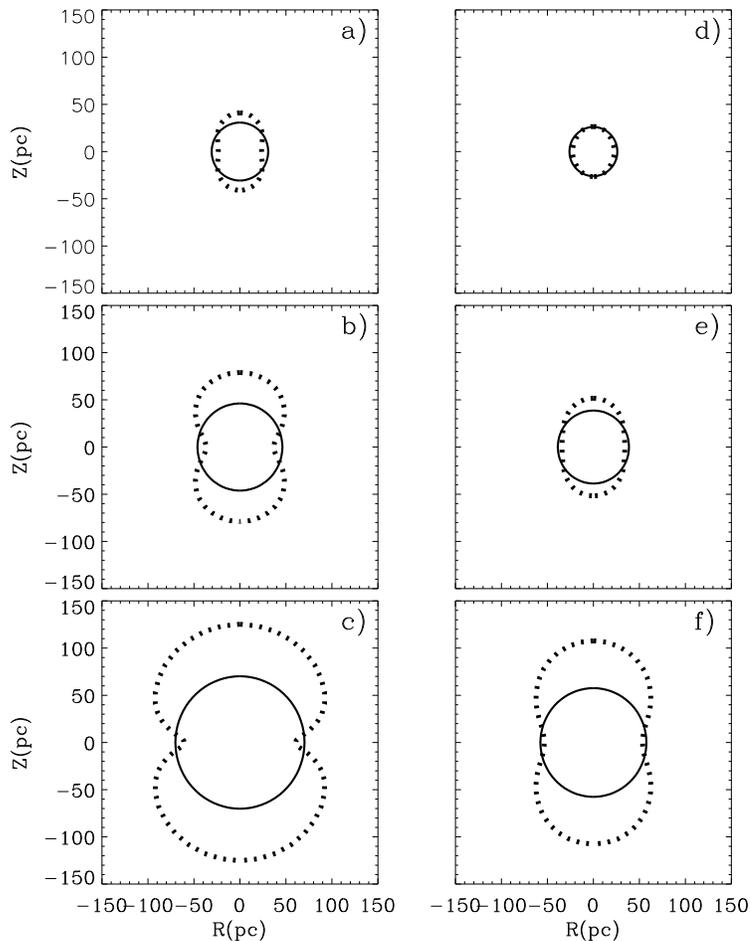}}
\caption{Superbubble shapes for models A1 and A2 (left panels), and B1 and
B2 (right panels). The panels show the evolution at $t$ = 1 Myr (a and d), 2
Myr (b and e), and 4 Myr (c and f). The solid lines represent the spherical
bubbles, A1 and B1, and the dotted lines correspond to models A2 and B2.}
\end{figure}
%---------------------------------------------------------------------
\clearpage
\begin{figure}
\centerline{\plotone{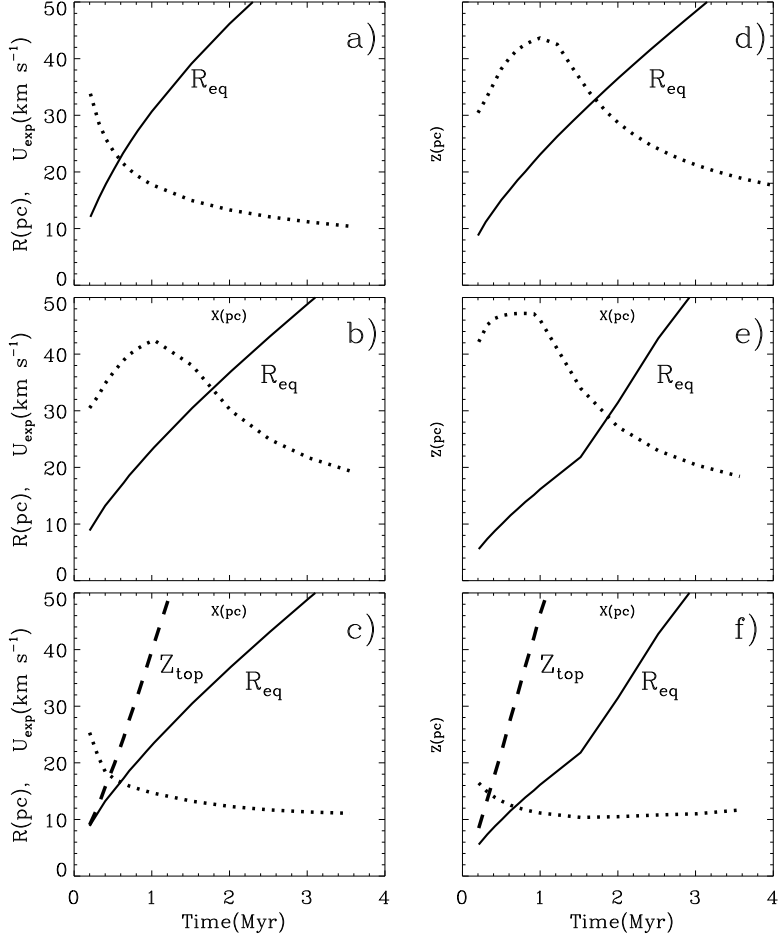}}
\caption{The expansion velocities and radii for models A. a) Spherical 
model A1. b) Model A2 as seen in a face-on galaxy. c) Model A2 as seen in an 
edge-on galaxy. d) Model A2 including shell fragmentation due to R-T 
instability. e) Model A3 for a face-on galaxy. f) Model A3 for an edge-on 
galaxy. The solid lines are the shell radii along the plane of the
galaxy, and the dashed lines are the top $z$-extensions. Dotted lines
for panels a, b, d and e are the top expansion velocities along the
$z$-axis. Dotted lines for panels c and f are the expansion
velocities at the bubble equator.} 
\end{figure}
%---------------------------------------------------------------------
\clearpage
\begin{figure}
\centerline{\plotone{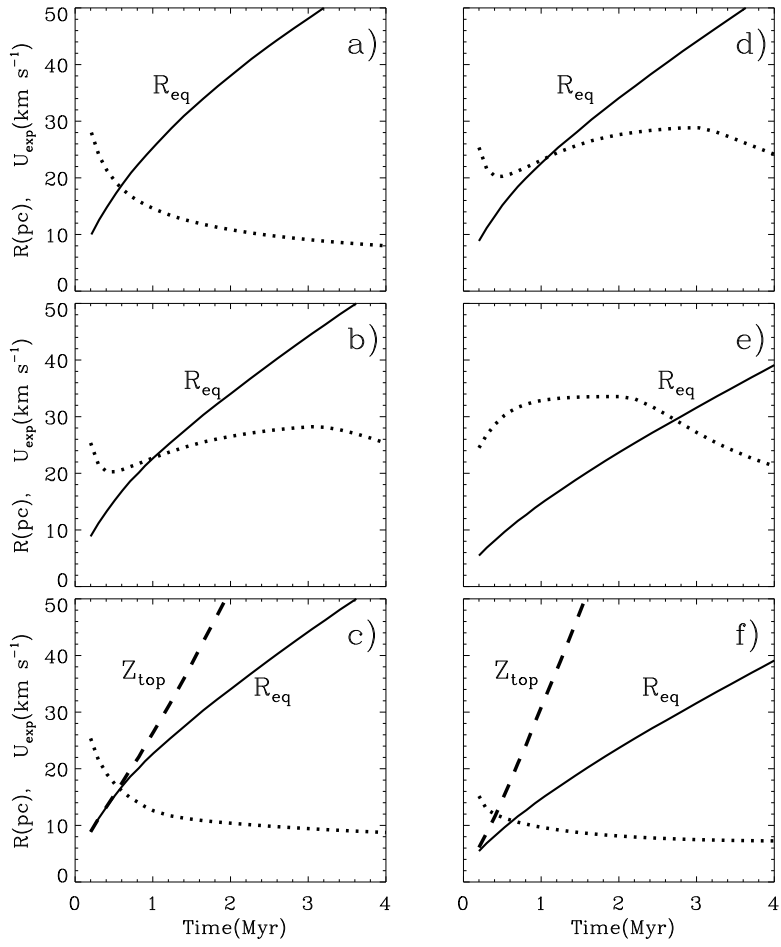}}
\caption{Same as Figure 3, but for models B.} 
\end{figure}
%---------------------------------------------------------------------
\clearpage
\begin{figure}
\centerline{\plotone{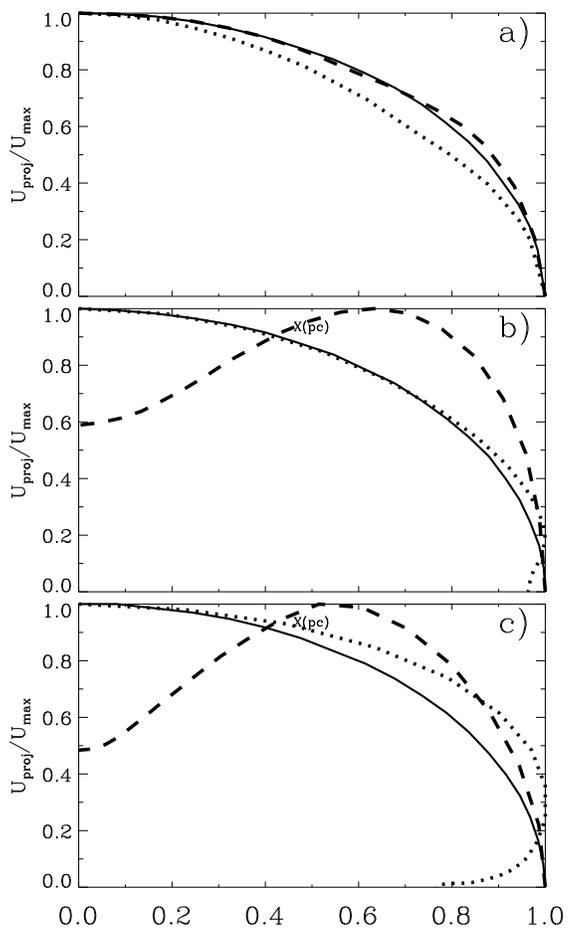}}
\caption{The normalized projected velocities as a function of the impact 
parameter $a$ for models A. a) $t = 2 \times 10^5$ yr, b) 1 Myr, and
c) 2 Myr. The solid lines present the spherical model A1, the dotted
lines show the face-on case A2, and the dashed lines are for the
edge-on case A2.} 
\end{figure}
%-------------------------------------------------------------
\clearpage
\begin{figure}
\centerline{\plotone{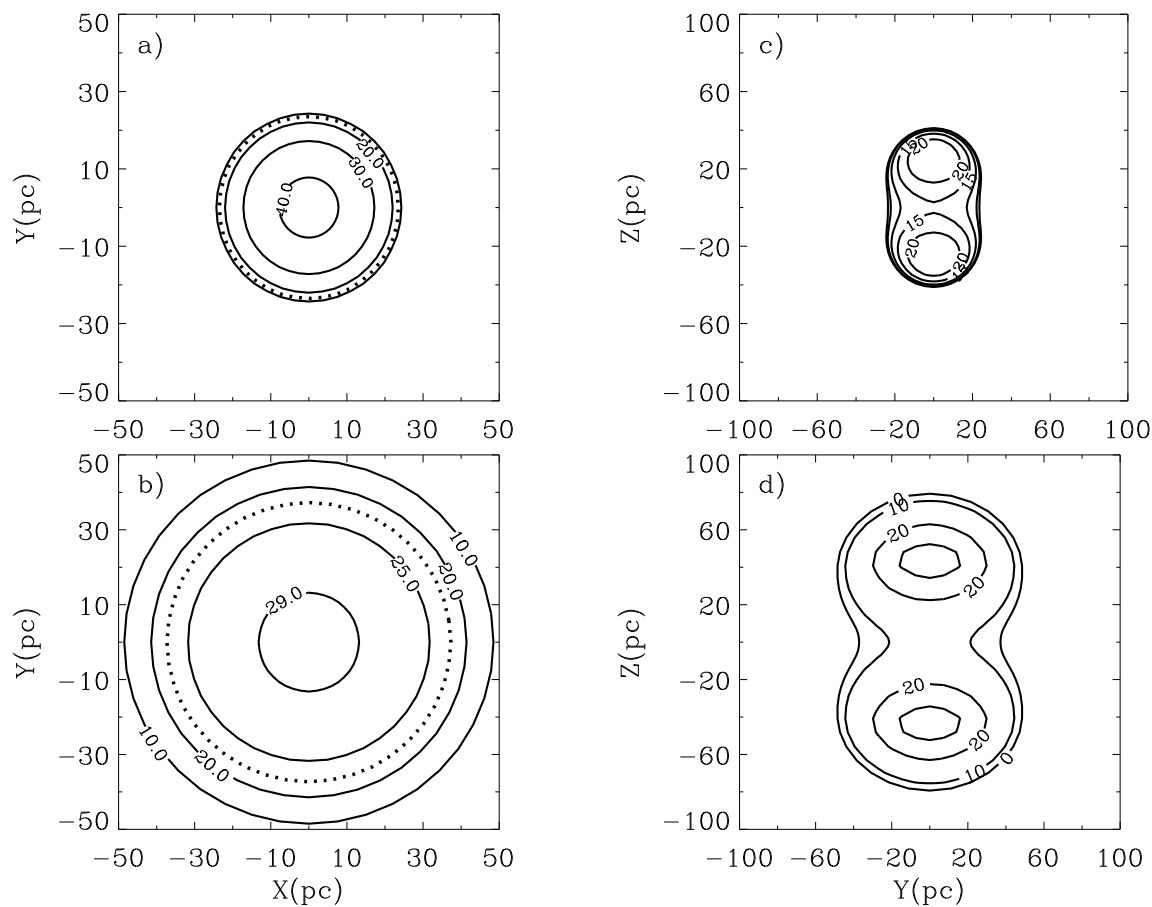}}
\caption{Isovelocity maps for model A2. Panels a) and b) show the results 
as seen in a face-on galaxy at 1 and 2 Myr, respectively. Panels c) and d) 
are the same for an edge-on galaxy. The dotted lines represent the zero 
velocity contours (\ie the bubble cross-section at the midplane of
the cloud).} 
\end{figure}
%-------------------------------------------------------------
\clearpage
\begin{figure}
\centerline{\plotone{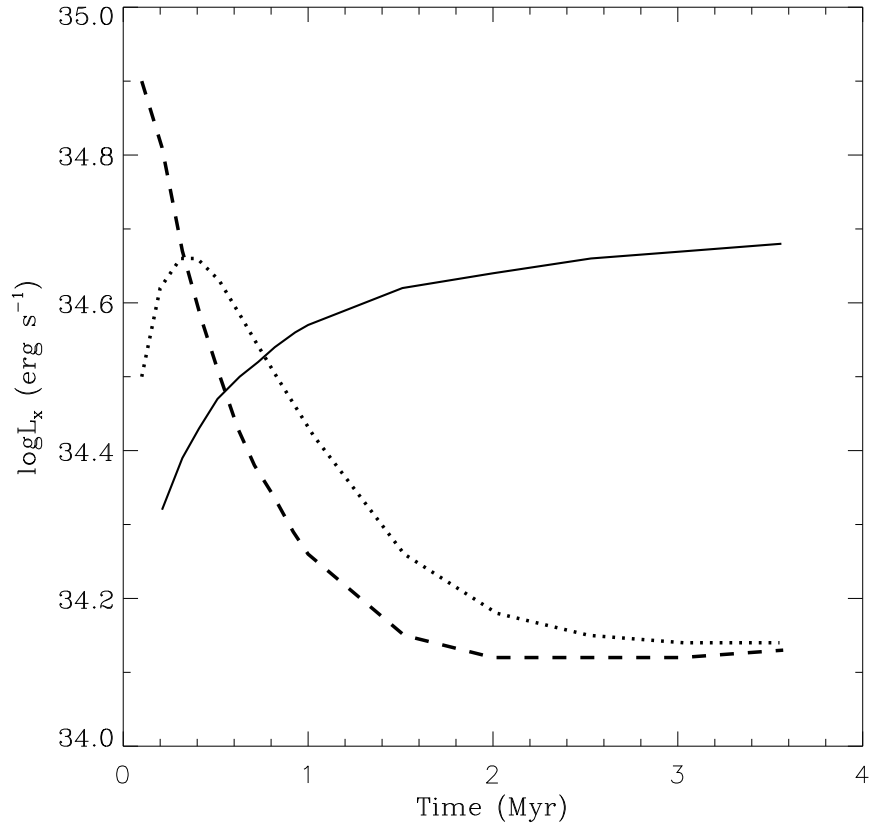}}
\caption{The X-ray luminosities for models A. The solid line presents
the spherical model A1, the dotted line is for model A2, and the
dashed line is for model A3.} 
\end{figure}

\end{document}